\documentclass[trackchanges,twocolumn]{aastex7}

\begin{document}

\title{A Chemical Mismatch Between Young Stars and Their Inner Disks}

\correspondingauthor{Diogo Souto}
\email{diogosouto@academico.ufs.br}

\author[0000-0002-7883-5425]{Diogo Souto}
\thanks{These authors contributed equally to this work.}
\affiliation{Departamento de F\'isica, Universidade Federal de Sergipe, Av. Marcelo Deda Chagas, S/N Cep 49.107-230, S\~ao Crist\'ov\~ao, SE, Brazil}
\email{diogosouto@academico.ufs.br}

\author[0000-0001-7962-1683]{Ilaria Pascucci}
\thanks{These authors contributed equally to this work.}
\affiliation{Lunar and Planetary Laboratory, The University of Arizona, Tucson, AZ 85721, USA}
\email{pascucci@arizona.edu}

\author[0000-0001-6476-0576]{Katia Cunha}
\affiliation{Steward Observatory, University of Arizona, 933 North Cherry Avenue, Tucson, AZ 85721-0065, USA}
\affiliation{Observatório Nacional/MCTIC, R. Gen. José Cristino, 77,  20921-400, Rio de Janeiro, Brazil}
\email{kcunha@arizona.edu}

\author[0000-0001-8401-4300]{Shubham Kanodia}
\affiliation{Carnegie Science Earth and Planets Laboratory, 5241 Broad Branch Road, NW, Washington, DC 20015, USA}
\email{skanodia@carnegiescience.edu}

\begin{abstract}
% No more than 250 words for Letters
We present the first stellar elemental abundance study for two very low-mass stars, similar in mass to TRAPPIST-1, in the $\sim5-10$\,Myr-old Upper-Sco association. Their mid-infrared JWST/MIRI spectra, like those of many very low-mass stars, are hydrocarbon-rich, indicating C/O ratios greater than unity in the inner disk gas inside their snowlines.  By fitting synthetic spectra to high-resolution APOGEE near-infrared stellar spectra, we show that, unlike their inner disks, both stars have solar C/O ratios. Their Fe, C, O, Mg, and Ca abundances are likewise consistent with solar values, placing them within the Galactic thin-disk population, as expected for nearby star-forming regions. This contrast between stellar and inner disk C/O ratios provides the first direct evidence that the inner disk's supersolar values are not inherited from the natal cloud but arise from disk processes. If these enhanced C/O ratios are primarily driven by inward drift of icy pebbles, there are major implications for disk evolution and planet formation, which we also discuss. 
\end{abstract}

%% Keywords should appear after the \end{abstract} command. 
%% The AAS Journals now uses Unified Astronomy Thesaurus (UAT) concepts:
%% https://astrothesaurus.org
%% You will be asked to selected these concepts during the submission process
%% but this old "keyword" functionality is maintained in case authors want
%% to include these concepts in their preprints.
%%
%% You can use the \uat command to link your UAT concepts back its source.

%% From the front matter, we move on to the body of the paper.
%% Sections are demarcated by \section and \subsection, respectively.
%% Observe the use of the LaTeX \label
%% command after the \subsection to give a symbolic KEY to the
%% subsection for cross-referencing in a \ref command.
%% You can use LaTeX's \ref and \label commands to keep track of
%% cross-references to sections, equations, tables, and figures.
%% That way, if you change the order of any elements, LaTeX will
%% automatically renumber them.

\section{Introduction}

The TRAPPIST-1 system is remarkable for its seven roughly Earth-sized planets orbiting a 0.1\,M$_\odot$ star, three of which lie within the habitable zone \citep[e.g.,][]{Agol2021PSJ.....2....1A}. It has become a cornerstone for studying planet formation and evolution around late M dwarfs, which are among the most common stars in the Galaxy \citep[e.g.,][]{Chabrier2005ASSL..327...41C}. Probing the chemical environments in which such planets form $-$  the inner disk regions inside the snowline $-$ is best achieved through infrared spectroscopy.

{\it Spitzer}/IRS observations of disks around very low-mass stars, similar in mass to TRAPPIST-1, revealed mid-infrared spectra distinct from those of young solar analogs, with weaker silicate emission features consistent with larger and more crystalline grains \citep{Apai2005Sci...310..834A,Pascucci2009ApJ...696..143P}. In the gas phase, these disks exhibit enhanced C$_2$H$_2$/HCN ratios and weak or absent H$_2$O lines, hinting at elevated C/O ratios inside the snowline \citep{Pascucci2009ApJ...696..143P,Pascucci2013ApJ...779..178P}. The sensitivity of JWST/MIRI has confirmed and extended this picture, revealing a rich hydrocarbon chemistry, including benzene detections, and firmly establishing that the inner disk gas is highly carbon-rich \citep[e.g.,][]{Tabone2023NatAs...7..805T,Arabhavi2024Sci...384.1086A,Grant2025A&A...702A.126G,Long2025ApJ...978L..30L}.

Several mechanisms have been proposed to explain these high gaseous C/O ratios. They include oxygen depletion from fast accretion onto the star of water vapor released from icy pebbles at the snowline \citep{Mah2023A&A...677L...7M,Sellek2025A&A...701A.239S}, or carbon enrichment via irreversible decomposition of solid-state organics at the soot line ($T \approx 400$ K, \citealt{Houge2025A&A...699A.227H}). 
It has also been proposed that reduced optical depth at the disk surface, due to overall larger grains, allows infrared spectroscopy to probe gas closer to the midplane $-$ potentially revealing a chemistry typical of disks around even sun-like stars but otherwise hidden from view \citep{Arabhavi2025,Jang2025arXiv250916004J}.

While the elevated inner disk C/O ratios are now well established and can be explained by several processes, the elemental abundances of the host stars, and in particular their C/O ratios, remain unknown. Here, we present the first such estimates using Apache Point Observatory Galactic Evolution Experiment (APOGEE) spectra.  APOGEE is a cryogenic, multi-fiber spectrograph \citep{Wilson2010} providing high-resolution (R $\approx$ 22,500) spectroscopy in the near infrared, specifically between 1.51--1.69 \micron .
Sect.~\ref{sec:apogee} describes the two stars selected from the larger sample of TRAPPIST-1-like stars with elevated inner disk C/O ratios and the APOGEE spectra utilized for the abundance analysis. Sect.\ref{sec:abundance} presents the methods and derived stellar parameters, and Sect.\ref{sec:results} shows that both stars have C/O ratios consistent with solar values. The main implications of this finding and future applications are summarized in Sect.\ref{sec:summary}.

\section{Sample and observational data} \label{sec:apogee}   
Determining elemental abundances of cool M dwarfs is notoriously more challenging than for hotter FGK dwarfs, mainly because of molecular absorption bands in their spectra. Recently, \cite{Souto2020,Souto2022ApJ...927..123S} used APOGEE spectra to derive M-dwarf abundances and benchmarked them against: (1) warm primary stars in wide binaries, which share the same original chemical composition as their M-dwarf companions, and (2) M dwarfs with precisely measured interferometric radii that provide accurate effective temperature constraints. The close agreement in elemental abundance ratios ([X/Fe]) between the M dwarfs and their warm primaries, differing on average by only $-0.05 \pm 0.03$\,dex, demonstrates that the derived M-dwarf abundances are consistent and precise within the observational uncertainties.

Elemental abundances of young, disk-bearing M stars are further complicated by veiling $-$ the weakening of spectral lines due to excess continuum emission from hot accretion shocks at the stellar surface (dominant at optical wavelengths, \citealt[e.g.,][]{Manara2017A&A...604A.127M}) and thermal emission from the hot inner disk (dominant at longer wavelengths, \citealt[e.g.,][]{Edwards2013ApJ...778..148E}). However, the late M-type stars whose inner disks exhibit elevated C/O ratios have low accretion rates and negligible infrared excess in the H band in comparison to sun-like stars \citep[e.g.,][]{Pascucci2013ApJ...779..178P}, which alleviates these effects. 
To identify suitable targets for abundance analyses, we cross-matched the sample of TRAPPIST-1-like stars with published infrared spectroscopy \citep{Pascucci2013ApJ...779..178P,Xie2023ApJ...959L..25X, Arabhavi2025,Long2025ApJ...978L..30L} with the APOGEE DR19 survey \citep{DR19} and found two sources in Upper Scorpius with available good quality spectra, 2MASS~J15582981-2310077 (J1558) and 2MASS~J16053215-1933159 (J1605). 

The two sources with APOGEE spectra not only belong to the same $\sim 5- 10$\,Myr-old  Upper Scorpius association \citep[e.g.,][]{Luhman2025AJ....170...19L} but also share the same spectral type (M4.5) and have nearly identical stellar luminosities and mass accretion rates \citep{Pascucci2013ApJ...779..178P,Arabhavi2025}. 
Their disks, however, show some differences. J1558's disk exhibits stronger infrared excess, has silicate emission features, and is detected at millimeter wavelengths with ALMA, whereas for J1605 only an upper limit is available \citep{Pascucci2013ApJ...779..178P,Carpenter2025ApJ...978..117C,Jang2025arXiv250916004J}. Although both inner disks have a gaseous C/O ratio greater than one \citep{Pascucci2013ApJ...779..178P,Long2025ApJ...978L..30L}, J1605 is significantly more carbon-rich than J1558, showing much stronger optically thick C$_2$H$_2$ emission that creates a pseudo-continuum, as well as the detection of benzene in its spectrum \citep{Tabone2023NatAs...7..805T,Arabhavi2025}. The main properties of these stars and their disks are summarized in Table~\ref{tab:abund}, and their spectral energy distributions with relevant portions of the JWST/MIRI spectra are shown in Figure~\ref{fig:spec}. 

% \section{Abundance analysis and Stellar parameters}
\section{Spectroscopic abundance analysis}
\label{sec:abundance}     

\subsection{Infrared spectra probing the disk atmosphere}

The infrared spectra of the disks discussed in this work were first obtained with Spitzer/IRS at a resolving power of $\sim 700$, and presented and analyzed in \cite{Pascucci2013ApJ...779..178P}. Molecular column densities for C-bearing molecules, and upper limits for water, were obtained by fitting the spectra with a slab of gas in local thermodynamic equilibrium (LTE) with molecular parameters from the HITRAN 2008 database (\citealt{Rothman2009JQSRT.110..533R}). The inferred column density ratios (C$_2$H$_2$/HCN vs. HCN/H$_2$O), when compared to predictions from disk chemical models by \cite{Najita2011ApJ...743..147N} across a range of C/O ratios, indicated C/O values greater than unity for the gas inside the snowline.

\begin{figure*}
    \centering
    \includegraphics[width=0.49\textwidth]{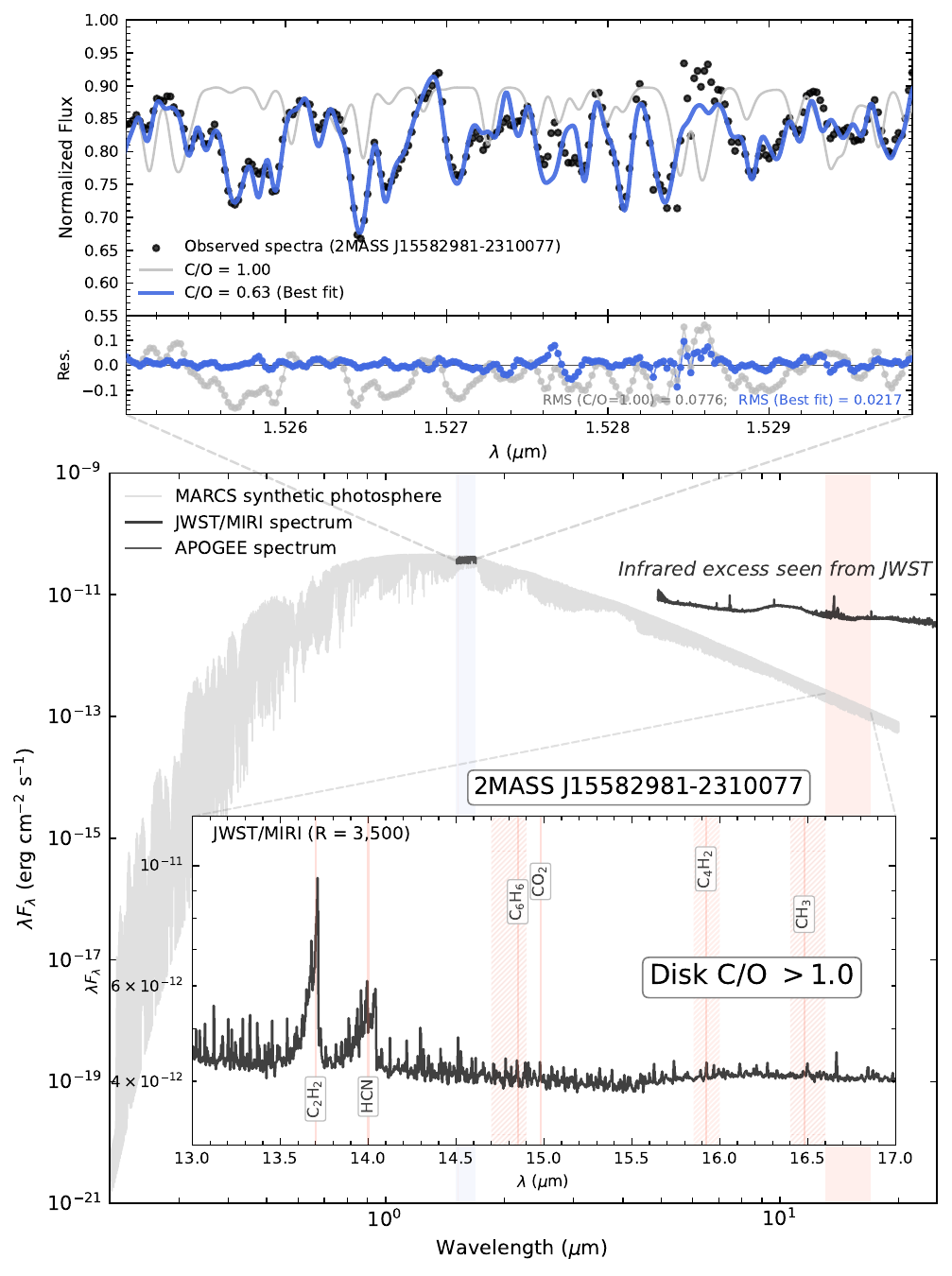}
    \includegraphics[width=0.49\textwidth]{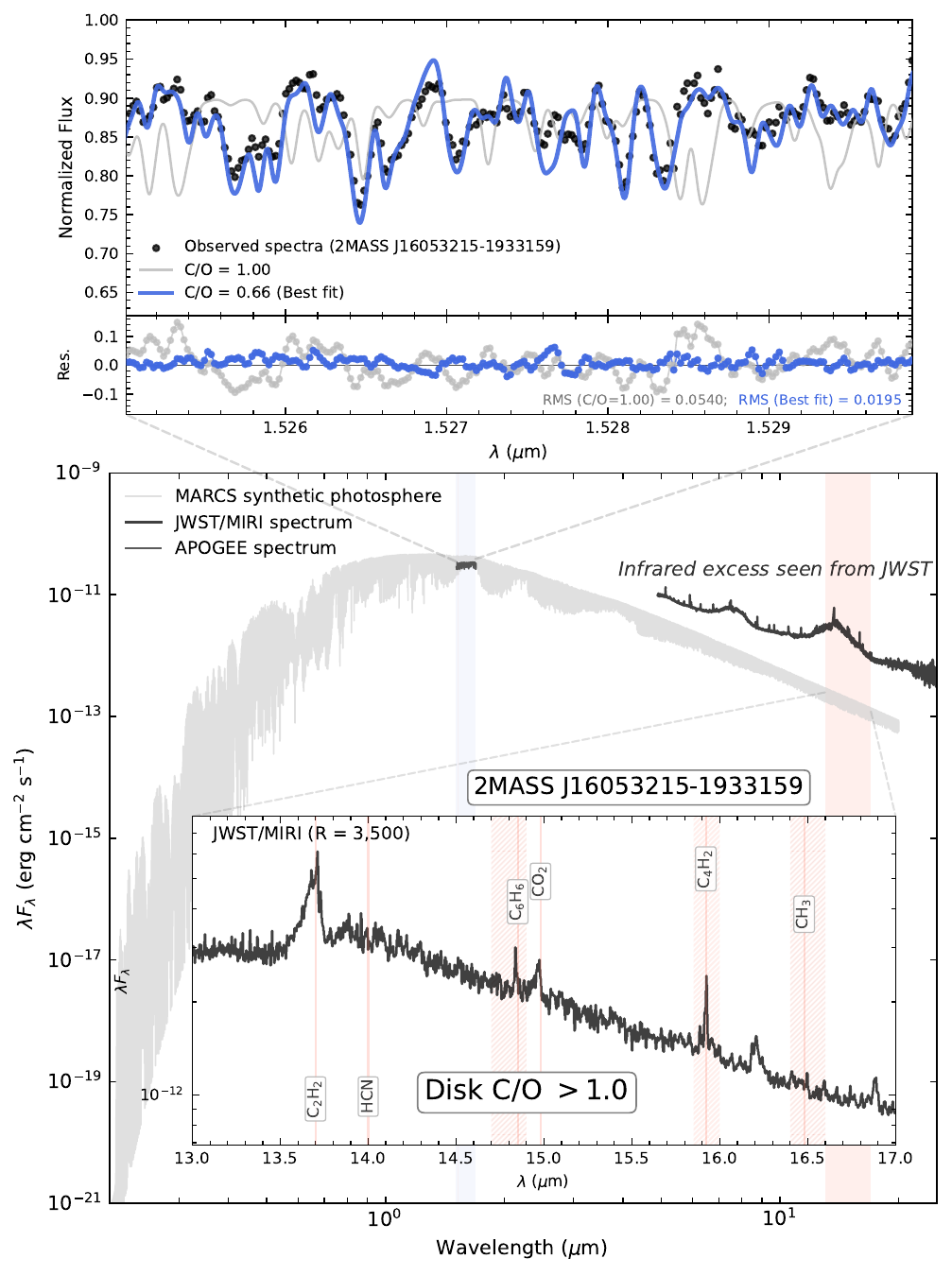}    
    \caption{Left and right panels show the same figure layout for \object{2MASS J15582981-2310077} and \object{2MASS J16053215-1933159}, respectively. The top panels show the high-resolution APOGEE spectra (black filled circles) and the best-fit synthetic spectra (in blue). Syntheses with C/O = 1 are also shown, as well as the residuals between syntheses and observed spectra. The bottom panels show the JWST/MIRI spectra together with the synthetic photospheric spectrum, while the lower inset highlights molecular features detected in the MIRI spectra.}
    \label{fig:spec}
\end{figure*}

The JWST Medium Resolution Spectrograph (MRS) spectra of the same disks were obtained as part of the European JWST Guaranteed Time Observing program 1282 (PI: Th. Henning) and published and analyzed in \cite{Tabone2023NatAs...7..805T,Arabhavi2025}. The spectra cover the $\sim 4.9$--$28.6\,\mu$m wavelength range at a resolving power of $\sim 2000-4000$, and we show in Figure~\ref{fig:spec} the portion that is rich in C-bearing molecules. For J1605, \cite{Tabone2023NatAs...7..805T} also carried out a detailed molecular column density retrieval using an approach similar to \cite{Pascucci2013ApJ...779..178P} but added a treatment of line overlap, which is particularly important for this source as it also shows an optically thick C$_2$H$_2$ continuum. The inferred C$_2$H$_2$/H$_2$O and C$_2$H$_2$/CO$_2$ ratios are orders of magnitude higher than those in the disks of young solar analogues, confirming a high C/O ratio in the inner gas disk of J1605.

Subsequently, the column density ratios of C$_2$H$_2$/CO$_2$ for J1558 and J1605 $-$ reported in \cite{Pascucci2013ApJ...779..178P} and \cite{Tabone2023NatAs...7..805T}, respectively $-$ were incorporated into a larger sample of disks around very low-mass stars with column densities inferred in a similar manner by \cite{Long2025ApJ...978L..30L}. Comparison with disk chemical models from \cite{Najita2011ApJ...743..147N} confirms that both sources in this study have inner gas disks with a C/O ratio greater than unity. Table \ref{tab:data} shows the disk C/O results from \cite{Long2025ApJ...978L..30L}.

\subsection{SDSS-V/APOGEE-2 spectra probing the stellar photosphere}

Individual abundances were derived by fitting theoretical synthetic spectra to the observed APOGEE spectra. 
The observed spectra are not corrected for extinction for two reasons. First, the extinction in the H band is small $-$ even for J1605 it is only $A_{\rm H} \approx 0.28$\,mag \citep[assuming a standard interstellar extinction law,][]{Mathis1990ARA&A..28...37M}. Second, the fitting is performed over a relatively narrow wavelength range and relies on continuum-normalized (non–flux-calibrated) spectra.
We generate synthetic spectra using one-dimensional plane-parallel MARCS atmospheric models (\citealt{Gustafsson2008}), which were interpolated when necessary to match the specific atmospheric parameters of each star. 
Atomic and molecular transitions for the spectral region between 1.5–1.7 \micron, numbering approximately 100,000 lines (excluding H$_2$O lines, which alone contribute nearly 2 million transitions), were taken from the APOGEE line list \citep{Smith2021AJ....161..254S}. 
These atmospheric models and line data were used as input to the radiative transfer calculations. We employed \texttt{Turbospectrum} combined with the \texttt{BACCHUS} wrapper to generate synthetic spectra (\citealt{AlvarezPlez1998}, \citealt{Plez2012}, \citealt{Masseron2016}). The \texttt{BACCHUS} interface facilitates direct comparisons between observed and computed spectra and provides $\chi^2$ metrics for each line, ensuring robustness in abundance determination analyses.

We determined the atmospheric parameters, effective temperature ($T_{\rm eff}$) and surface gravity (log $g$), for the studied M dwarfs by combining H$_{2}$O and OH lines, ensuring self-consistency in the derived oxygen abundances. Specifically, we varied $T_{\rm eff}$ from 2800 K to 3500 K in 100 K steps, noting that the OH lines are relatively insensitive to $T_{\rm eff}$, while the H$_{2}$O lines are more sensitive. Consequently, there is a unique $T_{\rm eff}$--A(O) combination that yields the same oxygen abundance from both sets of lines, which we adopt as our best-fit $T_{\rm eff}$.
Initially, we assume log $g$ = 4.50 for the $T_{\rm eff}$ determination. 
To refine our log $g$, we adopted the previously derived $T_{\rm eff}$ and we varied log $g$ from 4.1 to 5.2 dex in 0.1 dex increments, determining the oxygen abundance at each step. The obtained log $g$ is chosen where the oxygen abundance remains consistent across both the OH and H$_{2}$O lines, as we did for $T_{\rm eff}$. 
We then iterate this procedure until convergence in both $T_{\rm eff}$ and log $g$ is achieved. In practice, convergence is reached after two iterations.
We assumed the microturbulence parameter to be 1.00 km.s$^{-1}$ for all studied M dwarfs. Full details of this methodology are provided in \cite{Souto2020}.

Once the model atmospheres were selected for the two targets, we derived chemical abundances for %analyzed five 
the elements C, O, Mg, Ca, and Fe, using diagnostic lines % These elements were 
previously identified in the APOGEE spectra of M dwarfs (\citealt{Souto2017}; \citealt{Melo2024}). We note that we were unable to derive abundances for Na, Al, K, Ti, V, Cr, Mn, and Ni because their spectral lines are too weak at $T_{\rm eff}$ $\sim$3000 K. %in the spectral range.
Uncertainties in our derived abundances primarily arise from three sources: errors in the atmospheric parameters, the signal-to-noise ratio of the spectra, and the placement of the pseudo-continuum. We propagate these contributions in quadrature to estimate the final uncertainty in the abundance for each element. We adopted the abundance uncertainty estimates reported by \cite{Melo2024} as representative in this study.
Inferred stellar properties relevant to this study, including elemental abundances and their uncertainties, are summarized in Table~\ref{tab:abund}.

\begin{deluxetable*}{lrr}
\tablecaption{Stellar Parameters and Abundances \label{tab:abund}}
\tabletypesize{\scriptsize}
\label{tab:data}
\tablehead{
\colhead{Parameter} & \colhead{2Mass J1558} & \colhead{2Mass J1605}\\
\colhead{} & \colhead{2981$-$2310077} & \colhead{3215$-$1933159}
}
\startdata
\multicolumn{3}{c}{\textbf{Astrometric data}} \\
$^{\dagger}$RA & 239.62416 & 241.38391\\
$^{\dagger}$DEC & -23.168928 & -19.554546\\
$^{\dagger}$PMRA & $-12.641\pm0.058$ & $-10.408\pm0.068$\\
$^{\dagger}$PMDEC & -24.320$\pm$0.039 & -22.103$\pm$0.037\\
$^{\dagger}$Parallaxe & 7.0857 & 6.5647\\
% \multicolumn{3}{c}{\textbf{Photometric}} \\
% G (Gaia) & 15.626 & 16.245\\
% BP (Gaia) & 17.270 & 18.061\\
% RP (Gaia) & 14.356 & 14.937\\
% J (2Mass) & 12.297 & 12.587\\
% H (2Mass) & 11.670 & 11.894\\
% K$_{\rm S}$ (2Mass) & 11.300 & 11.358\\
% E(B-V) & 0.123$\pm$ 0.003 & 0.147$\pm$ 0.003 \\
\multicolumn{3}{c}{\textbf{Literature: star \& disk}} \\
$^{+}$SpTy    & M4.5  & M4.5 \\
$^{x}$log$L_*$ (L$_\odot$)  & -1.58 & -1.38 \\
$^{x}$$M_*$ (M$_\odot$)  & 0.11 & 0.14 \\
$^{\#+}$log$M_{\rm acc}$ (M$_\odot$/yr)  & -9.15 & -9.10\\
$^{\star}$$A_{\rm V}$  & 1.1 & 0.20\\
$^{\#}$$M_{\rm dust}$ (M$_\oplus$) & 1.50 & 0.24\\
$^{*}$C/O & 1.0$^{+0.2}_{-0.0}$ & 1.5$^{+0.5}_{-0.3}$ \\
\multicolumn{3}{c}{\textbf{This Work: spectroscopic results}} \\
SNR & 110 & 120 \\
$T_{\rm eff}$ (K) & 3073 $\pm$ 79 & 3000 $\pm$ 79 \\
$\log g$ (dex) & 4.35 $\pm$ 0.13 & 4.36 $\pm$ 0.13 \\
{[Fe/H]} & $-0.01 \pm 0.10$ (0.07) & $-0.07 \pm 0.10$ (0.08)\\
{[C/H]} & $0.01 \pm 0.07$ (0.04) & $-0.08 \pm 0.07$ (0.00)\\
{[C/Fe]} & 0.01 & -0.01 \\
{[O/H]} & $-0.01 \pm 0.05$ (0.03) & $-0.13 \pm 0.05$ (0.02)\\
{[O/Fe]} & $-0.02$ & $-0.06$ \\
{[C/O]} & 0.03$\pm$0.09 & 0.05$\pm$0.09 \\
C/O & 0.63$\pm$0.09 & 0.66$\pm$0.09 \\
{[Mg/H]} & 0.10 $\pm$ 0.12 (0.00) & -0.11 $\pm$ 0.12 (0.02)\\
{[Mg/Fe]} & 0.09 & -0.04 \\
{[Ca/H]} & $-0.04 \pm 0.07$ (0.02) & $-0.09 \pm 0.07$ (0.10)\\
{[Ca/Fe]} & $-0.05$ & -0.02 \\
\enddata
\tablecomments{The abundance uncertainties ($\pm$) represent the errors estimated following the prescription by \citet{Melo2024}, while values within the parentheses indicate the standard deviation of the mean abundances for the lines analyzed. Astrometric data are from \cite{Gaia2021}$^{\dagger}$, literature and star disk properties are from \cite{Pascucci2013ApJ...779..178P}$^{+}$ , \cite{HH2014ApJ...786...97H}$^{x}$ , \cite{Testi2022}$^{\#}$, \cite{Barenfeld2016ApJ...827..142B}$^{\star}$, and \cite{Long2025ApJ...978L..30L}$^{*}$.}
\end{deluxetable*}

\begin{figure*}
    \centering
    \includegraphics[width=0.8\textwidth]{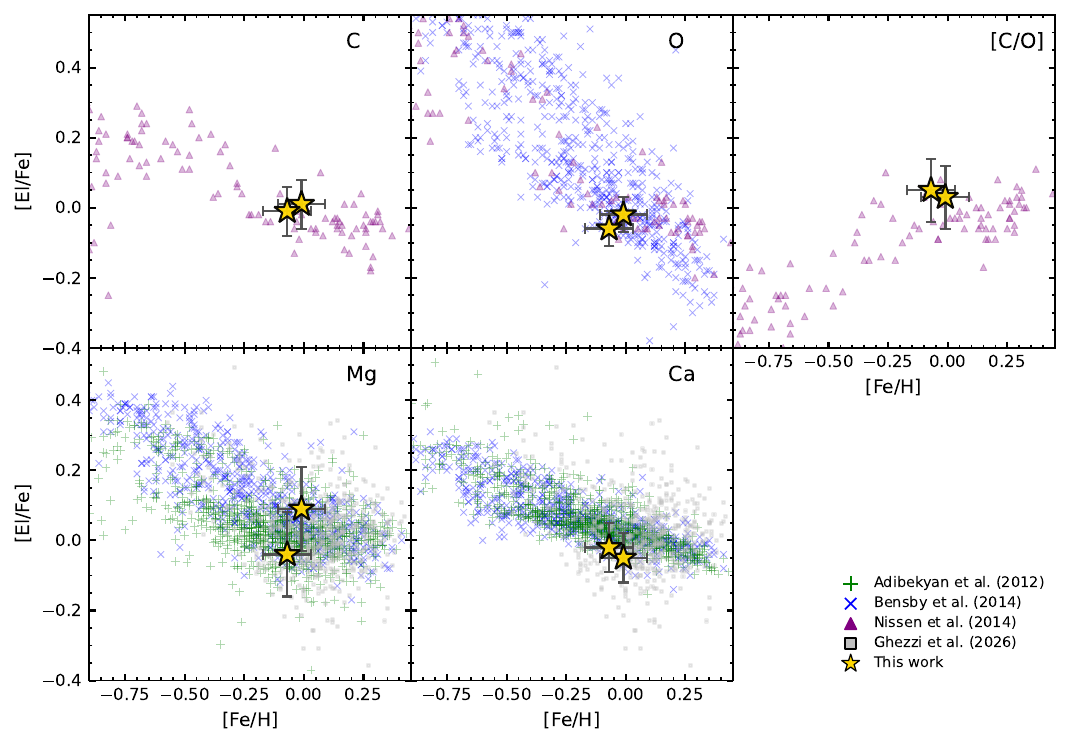}
    \caption{Abundance trends of [C/Fe], [O/Fe], [C/O], [Mg/Fe], and [Ca/Fe] \textit{vs.} [Fe/H]. The two M dwarfs from this work are shown as yellow stars, while the comparison Milky Way results are from \cite{Nissen2014A&A...568A..25N}, \cite{Adibekyan2012A&A...545A..32A}, \cite{Bensby2014A&A...562A..71B}, and \cite{Ghezzi2026ApJ...998..301G}.}
    \label{fig:abu}
\end{figure*}

\section{Results and discussion}
\label{sec:results} 

The effective temperatures derived from %APOGEE 
spectral analysis in this work 
of $T_{\rm eff}$= 3073 K for J1558, and $T_{\rm eff}$ = 3000 K for J1605 are fully consistent with independently derived SpTys from the literature \citep[see e.g., the spectral-type-temperature conversion in][]{HH2014ApJ...786...97H}. The star's surface gravity is as expected for a young, pre-main sequence M star \citep[e.g.,][]{Baraffe2015A&A...577A..42B}. The APOGEE spectra from this work were previously analyzed using a convolutional neural network pipeline (APOGEE net; \citealt{Sprague2022AJ....163..152S}), which obtained $T_{\rm eff}$ = 3190 K, log g = 4.55, and [Fe/H] = 0.013 for J1558, and  $T_{\rm eff}$ = 3218 K, log g = 4.37, [Fe/H] = 0.025 for J1605. Their effective temperatures are systematically hotter than ours by $\sim160$ K, while their metallicities are solar and consistent with our results.

In Figure \ref{fig:spec}, we show the MARCS synthetic photospheric flux (light gray) computed using the derived stellar $T_{\rm eff}$, together with the observed APOGEE spectrum (filled black circles; top panel) and the JWST/MIRI spectrum (dark gray). The near-infrared APOGEE region around $\lambda \sim 1.5$--$1.53\,\mu$m is shown in the upper inset, where the stellar C/O ratio is constrained from molecular absorption features. The best-fitting model (blue) is compared with a model assuming a super-stellar ${\rm C/O}=1$ in the photosphere (light gray). It is clear that a C/O ratio near 1 does not fit the observed APOGEE spectra. Our best-fitting model yields $\chi^2 \approx  1$, whereas a model assuming C/O = 1 results in $\chi^2 \approx  10$, corresponding to a fit that is roughly an order of magnitude worse. The residual diagram in the Figure also indicates this, showing that the RMS (root-mean-square) from our best fit is about three times smaller than that from assuming C/O = 1. 
The lower inset highlights the JWST/MIRI spectral region between $\sim 13$ and $17\,\mu$m, where emission features associated with hydrocarbons (e.g., C$_2$H$_2$, C$_4$H$_2$, and related species) are identified. These molecular emission features are consistent with carbon-rich disk chemistry (${\rm C/O}>1$), in contrast with the approximately solar stellar C/O ratios derived from the APOGEE spectra.
Importantly, the star's derived elemental abundances are roughly consistent with solar, and their C/O ratios of 0.63$\pm$0.09 and 0.66$\pm$0.09 for J1558 and J1605, respectively, are only slightly higher than the solar C/O ratio of 0.59 \citep{Asplund2021A&A...653A.141A}.

Figure \ref{fig:abu} shows Tinsley-Wallerstein diagrams of the determined [El/Fe] (C, O, Mg, Ca), and [C/O] ratios versus [Fe/H] for the two studied very low-mass stars (shown as yellow stars), compared with literature results from the optical for Milky Way disk stars from \cite{Nissen2014A&A...568A..25N}, \cite{Adibekyan2012A&A...545A..32A}, \cite{Bensby2014A&A...562A..71B}, and \cite{Ghezzi2026ApJ...998..301G}. Overall, the elemental abundance ratios for the two Upper Sco members fall, as expected, within the trends observed for the Galactic thin disk.

\section{Summary and Outlook} \label{sec:summary}

We presented the first retrieval of elemental abundances for two young very low-mass stars, similar in mass to TRAPPIST-1, whose mid-infrared spectra point to disk C/O ratios inside their snowlines greater than unity \citep{Pascucci2013ApJ...779..178P,Arabhavi2025}. In one of them,  the simple C$_2$H$_2$ hydrocarbon is so abundant as to create strong pseudo-continua around 7.5 and 14\,\micron\ \citep{Tabone2023NatAs...7..805T}. Our fit of theoretical synthetic spectra to the APOGEE stellar spectra demonstrates that the stars, unlike their inner disks, have a roughly solar C/O ratio.  The stellar elemental abundances of Fe, Mg, and Ca are also consistent with solar and place these stars squarely in the thin-disk population, as expected for stars in nearby star-forming regions \citep[e.g.,][]{Biazzo2012MNRAS.427.2905B,Spina2014A&A...568A...2S,Spina2017A&A...601A..70S}.

The implications of this finding are significant as it demonstrates that the elemental abundances of the star and the inner disk gas can differ. In a static disk picture, one would expect the gaseous C/O ratio inside the snowline $-$ probed with infrared spectroscopy $-$ to be close to solar as the main C and O carriers are in the gas phase \citep{O2011ApJ...743L..16O}. The elevated inner disk C/O ratios we observe from young TRAPPIST-1-like stars, therefore, imply that additional processes must act to substantially modify this basic expectation. If these processes are linked to disk dynamics, inward pebble migration \citep[leading either to O depletion and/or C enhancement in the inner disk][]{Mah2023A&A...677L...7M,Houge2025A&A...699A.227H,Sellek2025A&A...701A.239S}, they would have major implications for disk evolution and planet formation. First, disks around sun-like stars may also undergo a phase of elevated gaseous C/O ratios inside the snowline. Hints for such an evolution emerge from comparisons of JWST/MIRI spectra of young and old disk-bearing stars (Xie, Pascucci et al. submitted).
Second, planets forming from inward-migrating icy pebbles could attain bulk compositions very different from that of Earth. In particular, planets assembling between the water snowline and the inner organics-decomposition front could become highly carbon-rich ``soot''-like planets \citep[e.g.,][]{Li2026ApJ...997L..29L}. 
Finally, this type of analysis can be extended to other young stars in clusters to further compare stellar and disk chemistry.

\begin{acknowledgments}
We thank the referee for careful reading of the manuscript and helpful suggestions.
D.S. acknowledges support from the Foundation for Research and Technological Innovation Support of the State of Sergipe (FAPITEC/SE) and the National Council for Scientific and Technological Development (CNPq), under grant numbers 794017/2013 and 444372/2024-5. I.P. acknowledges partial support from NASA under agreement No. 80NSSC21K0593 for the program ``Alien Earths''.

Funding for the Sloan Digital Sky Survey V has been provided by the Alfred P. Sloan Foundation, the Heising-Simons Foundation, the National Science Foundation, and the Participating Institutions. SDSS acknowledges support and resources from the Center for High-Performance Computing at the University of Utah. SDSS telescopes are located at Apache Point Observatory, funded by the Astrophysical Research Consortium and operated by New Mexico State University, and at Las Campanas Observatory, operated by the Carnegie Institution for Science. The SDSS web site is \url{www.sdss.org}.
SDSS is managed by the Astrophysical Research Consortium for the Participating Institutions of the SDSS Collaboration, including Caltech, The Carnegie Institution for Science, Chilean National Time Allocation Committee (CNTAC) ratified researchers, The Flatiron Institute, the Gotham Participation Group, Harvard University, Heidelberg University, The Johns Hopkins University, L’Ecole polytechnique f\'{e}d\'{e}rale de Lausanne (EPFL), Leibniz-Institut f\"{u}r Astrophysik Potsdam (AIP), Max-Planck-Institut f\"{u}r Astronomie (MPIA Heidelberg), Max-Planck-Institut f\"{u}r Extraterrestrische Physik (MPE), Nanjing University, National Astronomical Observatories of China (NAOC), New Mexico State University, The Ohio State University, Pennsylvania State University, Smithsonian Astrophysical Observatory, Space Telescope Science Institute (STScI), the Stellar Astrophysics Participation Group, Universidad Nacional Aut\'{o}noma de M\'{e}xico, University of Arizona, University of Colorado Boulder, University of Illinois at Urbana-Champaign, University of Toronto, University of Utah, University of Virginia, Yale University, and Yunnan University.
The JWST data presented in this article were obtained from the Mikulski Archive for Space Telescopes (MAST) at the Space Telescope Science Institute. The specific observations analyzed can be accessed via \dataset[doi:10.17909/5d5a-es35]{doi:10.17909/5d5a-es35}.
\end{acknowledgments}

%\begin{contribution}
%%This section gives authors the space to recognize author contributions. The text inside this environment is NOT counted towards the total word quanta. At a minimum, manuscripts are expected to include this text:

%% But authors are expected to provide more specific details, e.g. 
%%
%%SC was responsible for writing and submitting the manuscript.
%%WWM came up with the initial research concept and edited the manuscript.
%%OTS obtained the funding and edited the manuscript.
%%EBF provided the formal analysis and validation. He also edited the manuscript.
%%GEH Supervised the undergraduates, wrote the software and administers the project github and Zenodo repositories.
%%
%% Authors can use the Contributor Role Taxonomy (CRediT) at
%% https://credit.niso.org
%% for ideas on how write a good statement tailored to their needs.

%\end{contribution}

%% To help institutions obtain information on the effectiveness of their 
%% telescopes the AAS Journals has created a group of keywords for telescope 
%% facilities.
%
%% Following the acknowledgments section, use the following syntax and the
%% \facility{} or \facilities{} macros to list the keywords of facilities used 
%% in the research for the paper.  Each keyword is check against the master 
%% list during copy editing.  Individual instruments can be provided in 
%% parentheses, after the keyword, but they are not verified.
\facilities{Sloan, Gaia}

%% Similar to \facility{}, there is the optional \software command to allow 
%% authors a place to specify which programs were used during the creation of 
%% the manuscript. Authors should list each code and include either a
%% citation or url to the code inside ()s when available.
\software{BACCHUS (\citealt{Masseron2016}), Turbospectrum (\citealt{AlvarezPlez1998}; \citealt{Plez2012}), Astropy \citep{2013A&A...558A..33A,2018AJ....156..123A, astropy:2022}, Numpy \citep{numpy}, Matplotlib \citep{matplotlib}, Scipy \citep{scipy}.
}

%% Appendix material should be preceded with a single \appendix command.
%% There should be a \section command for each appendix. Mark appendix
%% subsections with the same markup you use in the main body of the paper.
%%
%% Each Appendix (indicated with \section) will be lettered A, B, C, etc.
%% The equation counter will reset when it encounters the \appendix
%% command and will number appendix equations (A1), (A2), etc. The
%% Figure and Table counter will not reset.

\bibliography{sample7}{}

@ARTICLE{Grant2025A&A...702A.126G,
       author = {{Grant}, S.~L. and {Temmink}, M. and {van Dishoeck}, E.~F. and {Gasman}, D. and {Arabhavi}, A.~M. and {Tabone}, B. and {Henning}, T. and {Kamp}, I. and {Caratti o Garatti}, A. and {Christiaens}, V. and {Esteve}, P. and {G{\"u}del}, M. and {Jang}, H. and {Kaeufer}, T. and {Kurtovic}, N.~T. and {Morales-Calder{\'o}n}, M. and {Perotti}, G. and {Schwarz}, K. and {Sellek}, A.~D. and {Stapper}, L.~M. and {Vlasblom}, M. and {Waters}, L.~B.~F.~M.},
        title = "{MINDS: A transition from H$_{2}$O to C$_{2}$H$_{2}$ dominated disk spectra with decreasing stellar luminosity}",
      journal = {\aap},
         year = 2025,
        month = oct,
       volume = {702},
          eid = {A126},
        pages = {A126},
          doi = {10.1051/0004-6361/202555862},
archivePrefix = {arXiv},
       eprint = {2508.04692},
 primaryClass = {astro-ph.EP},
       adsurl = {https://ui.adsabs.harvard.edu/abs/2025A&A...702A.126G}
}

@ARTICLE{Testi2022,
       author = {{Testi}, L. and {Natta}, A. and {Manara}, C.~F. and {de Gregorio Monsalvo}, I. and {Lodato}, G. and {Lopez}, C. and {Muzic}, K. and {Pascucci}, I. and {Sanchis}, E. and {Miranda}, A. Santamaria and {Scholz}, A. and {De Simone}, M. and {Williams}, J.~P.},
        title = "{The protoplanetary disk population in the {\ensuremath{\rho}}-Ophiuchi region L1688 and the time evolution of Class II YSOs}",
      journal = {\aap},
         year = 2022,
        month = jul,
       volume = {663},
          eid = {A98},
        pages = {A98},
          doi = {10.1051/0004-6361/202141380},
archivePrefix = {arXiv},
       eprint = {2201.04079},
 primaryClass = {astro-ph.SR},
       adsurl = {https://ui.adsabs.harvard.edu/abs/2022A&A...663A..98T}
}

@ARTICLE{HH2014ApJ...786...97H,
       author = {{Herczeg}, Gregory J. and {Hillenbrand}, Lynne A.},
        title = "{An Optical Spectroscopic Study of T Tauri Stars. I. Photospheric Properties}",
      journal = {\apj},
         year = 2014,
        month = may,
       volume = {786},
       number = {2},
          eid = {97},
        pages = {97},
          doi = {10.1088/0004-637X/786/2/97},
archivePrefix = {arXiv},
       eprint = {1403.1675},
 primaryClass = {astro-ph.SR},
       adsurl = {https://ui.adsabs.harvard.edu/abs/2014ApJ...786...97H}
}

@ARTICLE{Xie2023ApJ...959L..25X,
       author = {{Xie}, Chengyan and {Pascucci}, Ilaria and {Long}, Feng and {Pontoppidan}, Klaus M. and {Banzatti}, Andrea and {Kalyaan}, Anusha and {Salyk}, Colette and {Liu}, Yao and {Najita}, Joan R. and {Pinilla}, Paola and {Arulanantham}, Nicole and {Herczeg}, Gregory J. and {Carr}, John and {Bergin}, Edwin A. and {Ballering}, Nicholas P. and {Krijt}, Sebastiaan and {Blake}, Geoffrey A. and {Zhang}, Ke and {{\"O}berg}, Karin I. and {Green}, Joel D. and {Jdiscs Collaboration}},
        title = "{Water-rich Disks around Late M Stars Unveiled: Exploring the Remarkable Case of Sz 114}",
      journal = {\apjl},
         year = 2023,
        month = dec,
       volume = {959},
       number = {2},
          eid = {L25},
        pages = {L25},
          doi = {10.3847/2041-8213/ad0ed9},
archivePrefix = {arXiv},
       eprint = {2310.13205},
 primaryClass = {astro-ph.EP},
       adsurl = {https://ui.adsabs.harvard.edu/abs/2023ApJ...959L..25X}
}

@ARTICLE{Mathis1990ARA&A..28...37M,
       author = {{Mathis}, John S.},
        title = "{Interstellar dust and extinction.}",
      journal = {\araa},
         year = 1990,
        month = jan,
       volume = {28},
        pages = {37-70},
          doi = {10.1146/annurev.aa.28.090190.000345},
       adsurl = {https://ui.adsabs.harvard.edu/abs/1990ARA&A..28...37M}
}

@ARTICLE{Baraffe2015A&A...577A..42B,
       author = {{Baraffe}, Isabelle and {Homeier}, Derek and {Allard}, France and {Chabrier}, Gilles},
        title = "{New evolutionary models for pre-main sequence and main sequence low-mass stars down to the hydrogen-burning limit}",
      journal = {\aap},
         year = 2015,
        month = may,
       volume = {577},
          eid = {A42},
        pages = {A42},
          doi = {10.1051/0004-6361/201425481},
archivePrefix = {arXiv},
       eprint = {1503.04107},
 primaryClass = {astro-ph.SR},
       adsurl = {https://ui.adsabs.harvard.edu/abs/2015A&A...577A..42B}
}

@INPROCEEDINGS{Chabrier2005ASSL..327...41C,
       author = {{Chabrier}, Gilles},
        title = "{The Initial Mass Function: From Salpeter 1955 to 2005}",
    booktitle = {The Initial Mass Function 50 Years Later},
         year = 2005,
       editor = {{Corbelli}, E. and {Palla}, F. and {Zinnecker}, H.},
       series = {Astrophysics and Space Science Library},
       volume = {327},
        month = jan,
        pages = {41},
          doi = {10.1007/978-1-4020-3407-7_5},
archivePrefix = {arXiv},
       eprint = {astro-ph/0409465},
 primaryClass = {astro-ph},
       adsurl = {https://ui.adsabs.harvard.edu/abs/2005ASSL..327...41C}
}

@ARTICLE{Li2026ApJ...997L..29L,
       author = {{Li}, Jie and {Bergin}, Edwin A. and {Hirschmann}, Marc M. and {Blake}, Geoffrey A. and {Ciesla}, Fred J. and {Kempton}, Eliza M.-R.},
        title = "{Soot Planets Instead of Water Worlds}",
      journal = {\apjl},
         year = 2026,
        month = jan,
       volume = {997},
       number = {1},
          eid = {L29},
        pages = {L29},
          doi = {10.3847/2041-8213/ae29a6},
archivePrefix = {arXiv},
       eprint = {2508.16781},
 primaryClass = {astro-ph.EP},
       adsurl = {https://ui.adsabs.harvard.edu/abs/2026ApJ...997L..29L}
}

@ARTICLE{O2011ApJ...743L..16O,
       author = {{{\"O}berg}, Karin I. and {Murray-Clay}, Ruth and {Bergin}, Edwin A.},
        title = "{The Effects of Snowlines on C/O in Planetary Atmospheres}",
      journal = {\apjl},
         year = 2011,
        month = dec,
       volume = {743},
       number = {1},
          eid = {L16},
        pages = {L16},
          doi = {10.1088/2041-8205/743/1/L16},
archivePrefix = {arXiv},
       eprint = {1110.5567},
 primaryClass = {astro-ph.GA},
       adsurl = {https://ui.adsabs.harvard.edu/abs/2011ApJ...743L..16O}
}

@ARTICLE{Carpenter2025ApJ...978..117C,
       author = {{Carpenter}, John M. and {Esplin}, Taran L. and {Luhman}, Kevin L. and {Mamajek}, Eric E. and {Andrews}, Sean M.},
        title = "{Extending the ALMA Census of Circumstellar Disks in the Upper Scorpius OB Association}",
      journal = {\apj},
         year = 2025,
        month = jan,
       volume = {978},
       number = {1},
          eid = {117},
        pages = {117},
          doi = {10.3847/1538-4357/ad8ebc},
archivePrefix = {arXiv},
       eprint = {2410.21598},
 primaryClass = {astro-ph.SR},
       adsurl = {https://ui.adsabs.harvard.edu/abs/2025ApJ...978..117C}
}

@ARTICLE{Luhman2025AJ....170...19L,
       author = {{Luhman}, K.~L.},
        title = "{The Initial Mass Function of Stars and Brown Dwarfs in the Upper Sco Association}",
      journal = {\aj},
         year = 2025,
        month = jul,
       volume = {170},
       number = {1},
          eid = {19},
        pages = {19},
          doi = {10.3847/1538-3881/add68c},
archivePrefix = {arXiv},
       eprint = {2505.21747},
 primaryClass = {astro-ph.SR},
       adsurl = {https://ui.adsabs.harvard.edu/abs/2025AJ....170...19L}
}

@ARTICLE{Manara2017A&A...604A.127M,
       author = {{Manara}, C.~F. and {Testi}, L. and {Herczeg}, G.~J. and {Pascucci}, I. and {Alcal{\'a}}, J.~M. and {Natta}, A. and {Antoniucci}, S. and {Fedele}, D. and {Mulders}, G.~D. and {Henning}, T. and {Mohanty}, S. and {Prusti}, T. and {Rigliaco}, E.},
        title = "{X-shooter study of accretion in Chamaeleon I. II. A steeper increase of accretion with stellar mass for very low-mass stars?}",
      journal = {\aap},
         year = 2017,
        month = aug,
       volume = {604},
          eid = {A127},
        pages = {A127},
          doi = {10.1051/0004-6361/201630147},
archivePrefix = {arXiv},
       eprint = {1704.02842},
 primaryClass = {astro-ph.SR},
       adsurl = {https://ui.adsabs.harvard.edu/abs/2017A&A...604A.127M}
}

@ARTICLE{Edwards2013ApJ...778..148E,
       author = {{Edwards}, Suzan and {Kwan}, John and {Fischer}, William and {Hillenbrand}, Lynne and {Finn}, Kimberly and {Fedorenko}, Kristina and {Feng}, Wanda},
        title = "{Interpreting Near-infrared Hydrogen Line Ratios in T Tauri Stars}",
      journal = {\apj},
         year = 2013,
        month = dec,
       volume = {778},
       number = {2},
          eid = {148},
        pages = {148},
          doi = {10.1088/0004-637X/778/2/148},
archivePrefix = {arXiv},
       eprint = {1309.4449},
 primaryClass = {astro-ph.SR},
       adsurl = {https://ui.adsabs.harvard.edu/abs/2013ApJ...778..148E}
}

@ARTICLE{Souto2022ApJ...927..123S,
       author = {{Souto}, Diogo and {Cunha}, Katia and {Smith}, Verne V. and {Allende Prieto}, C. and {Covey}, Kevin and {Garc{\'\i}a-Hern{\'a}ndez}, D.~A. and {Holtzman}, Jon A. and {J{\"o}nsson}, Henrik and {Mahadevan}, Suvrath and {Majewski}, Steven R. and {Masseron}, Thomas and {Pinsonneault}, Marc and {Schneider}, Donald P. and {Shetrone}, Matthew and {Stassun}, Keivan G. and {Terrien}, Ryan and {Zamora}, Olga and {Stringfellow}, Guy S. and {Lane}, Richard R. and {Nitschelm}, Christian and {Rojas-Ayala}, B{\'a}rbara},
        title = "{Detailed Chemical Abundances for a Benchmark Sample of M Dwarfs from the APOGEE Survey}",
      journal = {\apj},
         year = 2022,
        month = mar,
       volume = {927},
       number = {1},
          eid = {123},
        pages = {123},
          doi = {10.3847/1538-4357/ac4891},
archivePrefix = {arXiv},
       eprint = {2201.00891},
 primaryClass = {astro-ph.SR},
       adsurl = {https://ui.adsabs.harvard.edu/abs/2022ApJ...927..123S}
}

@ARTICLE{Jang2025arXiv250916004J,
       author = {{Jang}, Hyerin and {Arabhavi}, Aditya M. and {Kaeufer}, Till and {Waters}, Rens and {Kamp}, Inga and {Henning}, Thomas and {Garatti}, Alessio Caratti o and {van Dishoeck}, Ewine F. and {Perotti}, Giulia and {Kanwar}, Jayatee and {G{\"u}del}, Manuel and {Morales-Calder{\'o}n}, Maria and {Grant}, Sierra L. and {Christiaens}, Valentin},
        title = "{MINDS: The very low-mass star and brown dwarf sample II. Probing disk settling, dust properties, and dust-gas interplay with JWST/MIRI}",
      journal = {arXiv e-prints},
         year = 2025,
        month = sep,
          eid = {arXiv:2509.16004},
        pages = {arXiv:2509.16004},
          doi = {10.48550/arXiv.2509.16004},
archivePrefix = {arXiv},
       eprint = {2509.16004},
 primaryClass = {astro-ph.EP},
       adsurl = {https://ui.adsabs.harvard.edu/abs/2025arXiv250916004J}
}

@ARTICLE{Houge2025A&A...699A.227H,
       author = {{Houge}, Adrien and {Johansen}, Anders and {Bergin}, Edwin and {Ciesla}, Fred J. and {Bitsch}, Bertram and {Lambrechts}, Michiel and {Henning}, Thomas and {Perotti}, Giulia},
        title = "{Burned to ashes: How the thermal decomposition of refractory organics in the inner protoplanetary disc impacts the gas-phase C/O ratio}",
      journal = {\aap},
         year = 2025,
        month = jul,
       volume = {699},
          eid = {A227},
        pages = {A227},
          doi = {10.1051/0004-6361/202555164},
archivePrefix = {arXiv},
       eprint = {2505.20427},
 primaryClass = {astro-ph.EP},
       adsurl = {https://ui.adsabs.harvard.edu/abs/2025A&A...699A.227H}
}

@ARTICLE{Sellek2025A&A...701A.239S,
       author = {{Sellek}, Andrew D. and {van Dishoeck}, Ewine F.},
        title = "{Chemical transformation of CO in evolving protoplanetary discs across stellar masses: A route to C-rich inner regions}",
      journal = {\aap},
         year = 2025,
        month = sep,
       volume = {701},
          eid = {A239},
        pages = {A239},
          doi = {10.1051/0004-6361/202555195},
archivePrefix = {arXiv},
       eprint = {2507.11631},
 primaryClass = {astro-ph.EP},
       adsurl = {https://ui.adsabs.harvard.edu/abs/2025A&A...701A.239S}
}

@ARTICLE{Mah2023A&A...677L...7M,
       author = {{Mah}, Jingyi and {Bitsch}, Bertram and {Pascucci}, Ilaria and {Henning}, Thomas},
        title = "{Close-in ice lines and the super-stellar C/O ratio in discs around very low-mass stars}",
      journal = {\aap},
         year = 2023,
        month = sep,
       volume = {677},
          eid = {L7},
        pages = {L7},
          doi = {10.1051/0004-6361/202347169},
archivePrefix = {arXiv},
       eprint = {2308.15128},
 primaryClass = {astro-ph.EP},
       adsurl = {https://ui.adsabs.harvard.edu/abs/2023A&A...677L...7M}
}

@ARTICLE{Long2025ApJ...978L..30L,
       author = {{Long}, Feng and {Pascucci}, Ilaria and {Houge}, Adrien and {Banzatti}, Andrea and {Pontoppidan}, Klaus M. and {Najita}, Joan and {Krijt}, Sebastiaan and {Xie}, Chengyan and {Williams}, Joe and {Herczeg}, Gregory J. and {Andrews}, Sean M. and {Bergin}, Edwin and {Blake}, Geoffrey A. and {Colmenares}, Mar{\'\i}a Jos{\'e} and {Harsono}, Daniel and {Romero-Mirza}, Carlos E. and {Li}, Rixin and {Lu}, Cicero X. and {Pinilla}, Paola and {Wilner}, David J. and {Vioque}, Miguel and {Zhang}, Ke and {JDISCS Collaboration}},
        title = "{The First JWST View of a 30-Myr-old Protoplanetary Disk Reveals a Late-stage Carbon-rich Phase}",
      journal = {\apjl},
         year = 2025,
        month = jan,
       volume = {978},
       number = {2},
          eid = {L30},
        pages = {L30},
          doi = {10.3847/2041-8213/ad99d2},
archivePrefix = {arXiv},
       eprint = {2412.05535},
 primaryClass = {astro-ph.EP},
       adsurl = {https://ui.adsabs.harvard.edu/abs/2025ApJ...978L..30L}
}

@ARTICLE{Arabhavi2025,
       author = {{Arabhavi}, A.~M. and {Kamp}, I. and {Henning}, Th. and {van Dishoeck}, E.~F. and {Jang}, H. and {Waters}, L.~B.~F.~M. and {Christiaens}, V. and {Gasman}, D. and {Pascucci}, I. and {Perotti}, G. and {Grant}, S.~L. and {G{\"u}del}, M. and {Lagage}, P.-O. and {Barrado}, D. and {Caratti o Garatti}, A. and {Lahuis}, F. and {Kaeufer}, T. and {Kanwar}, J. and {Morales-Calder{\'o}n}, M. and {Schwarz}, K. and {Sellek}, A.~D. and {Tabone}, B. and {Temmink}, M. and {Vlasblom}, M. and {Patapis}, P.},
        title = "{MINDS: The very low-mass star and brown dwarf sample: Detections and trends in the inner disk gas}",
      journal = {\aap},
         year = 2025,
        month = jul,
       volume = {699},
          eid = {A194},
        pages = {A194},
          doi = {10.1051/0004-6361/202554109},
archivePrefix = {arXiv},
       eprint = {2506.02748},
 primaryClass = {astro-ph.EP},
       adsurl = {https://ui.adsabs.harvard.edu/abs/2025A&A...699A.194A}
}

@ARTICLE{Tabone2023NatAs...7..805T,
       author = {{Tabone}, B. and {Bettoni}, G. and {van Dishoeck}, E.~F. and {Arabhavi}, A.~M. and {Grant}, S. and {Gasman}, D. and {Henning}, Th. and {Kamp}, I. and {G{\"u}del}, M. and {Lagage}, P.~O. and {Ray}, T. and {Vandenbussche}, B. and {Abergel}, A. and {Absil}, O. and {Argyriou}, I. and {Barrado}, D. and {Boccaletti}, A. and {Bouwman}, J. and {Caratti o Garatti}, A. and {Geers}, V. and {Glauser}, A.~M. and {Justannont}, K. and {Lahuis}, F. and {Mueller}, M. and {Nehm{\'e}}, C. and {Olofsson}, G. and {Pantin}, E. and {Scheithauer}, S. and {Waelkens}, C. and {Waters}, L.~B.~F.~M. and {Black}, J.~H. and {Christiaens}, V. and {Guadarrama}, R. and {Morales-Calder{\'o}n}, M. and {Jang}, H. and {Kanwar}, J. and {Pawellek}, N. and {Perotti}, G. and {Perrin}, A. and {Rodgers-Lee}, D. and {Samland}, M. and {Schreiber}, J. and {Schwarz}, K. and {Colina}, L. and {{\"O}stlin}, G. and {Wright}, G.},
        title = "{A rich hydrocarbon chemistry and high C to O ratio in the inner disk around a very low-mass star}",
      journal = {Nature Astronomy},
         year = 2023,
        month = jul,
       volume = {7},
        pages = {805-814},
          doi = {10.1038/s41550-023-01965-3},
archivePrefix = {arXiv},
       eprint = {2304.05954},
 primaryClass = {astro-ph.EP},
       adsurl = {https://ui.adsabs.harvard.edu/abs/2023NatAs...7..805T}
}

@ARTICLE{Arabhavi2024Sci...384.1086A,
       author = {{Arabhavi}, A.~M. and {Kamp}, I. and {Henning}, Th. and {van Dishoeck}, E.~F. and {Christiaens}, V. and {Gasman}, D. and {Perrin}, A. and {G{\"u}del}, M. and {Tabone}, B. and {Kanwar}, J. and {Waters}, L.~B.~F.~M. and {Pascucci}, I. and {Samland}, M. and {Perotti}, G. and {Bettoni}, G. and {Grant}, S.~L. and {Lagage}, P.~O. and {Ray}, T.~P. and {Vandenbussche}, B. and {Absil}, O. and {Argyriou}, I. and {Barrado}, D. and {Boccaletti}, A. and {Bouwman}, J. and {Caratti o Garatti}, A. and {Glauser}, A.~M. and {Lahuis}, F. and {Mueller}, M. and {Olofsson}, G. and {Pantin}, E. and {Scheithauer}, S. and {Morales-Calder{\'o}n}, M. and {Franceschi}, R. and {Jang}, H. and {Pawellek}, N. and {Rodgers-Lee}, D. and {Schreiber}, J. and {Schwarz}, K. and {Temmink}, M. and {Vlasblom}, M. and {Wright}, G. and {Colina}, L. and {{\"O}stlin}, G.},
        title = "{Abundant hydrocarbons in the disk around a very-low-mass star}",
      journal = {Science},
         year = 2024,
        month = jun,
       volume = {384},
       number = {6700},
        pages = {1086-1090},
          doi = {10.1126/science.adi8147},
archivePrefix = {arXiv},
       eprint = {2406.14293},
 primaryClass = {astro-ph.EP},
       adsurl = {https://ui.adsabs.harvard.edu/abs/2024Sci...384.1086A}
}

@ARTICLE{Pascucci2013ApJ...779..178P,
       author = {{Pascucci}, I. and {Herczeg}, G. and {Carr}, J.~S. and {Bruderer}, S.},
        title = "{The Atomic and Molecular Content of Disks around Very Low-mass Stars and Brown Dwarfs}",
      journal = {\apj},
         year = 2013,
        month = dec,
       volume = {779},
       number = {2},
          eid = {178},
        pages = {178},
          doi = {10.1088/0004-637X/779/2/178},
archivePrefix = {arXiv},
       eprint = {1311.1228},
 primaryClass = {astro-ph.EP},
       adsurl = {https://ui.adsabs.harvard.edu/abs/2013ApJ...779..178P}
}

@ARTICLE{Pascucci2009ApJ...696..143P,
       author = {{Pascucci}, I. and {Apai}, D. and {Luhman}, K. and {Henning}, Th. and {Bouwman}, J. and {Meyer}, M.~R. and {Lahuis}, F. and {Natta}, A.},
        title = "{The Different Evolution of Gas and Dust in Disks around Sun-Like and Cool Stars}",
      journal = {\apj},
         year = 2009,
        month = may,
       volume = {696},
       number = {1},
        pages = {143-159},
          doi = {10.1088/0004-637X/696/1/143},
archivePrefix = {arXiv},
       eprint = {0810.2552},
 primaryClass = {astro-ph},
       adsurl = {https://ui.adsabs.harvard.edu/abs/2009ApJ...696..143P}
}

@ARTICLE{Apai2005Sci...310..834A,
       author = {{Apai}, D{\'a}niel and {Pascucci}, Ilaria and {Bouwman}, Jeroen and {Natta}, Antonella and {Henning}, Thomas and {Dullemond}, Cornelis P.},
        title = "{The Onset of Planet Formation in Brown Dwarf Disks}",
      journal = {Science},
         year = 2005,
        month = nov,
       volume = {310},
       number = {5749},
        pages = {834-836},
          doi = {10.1126/science.1118042},
archivePrefix = {arXiv},
       eprint = {astro-ph/0511420},
 primaryClass = {astro-ph},
       adsurl = {https://ui.adsabs.harvard.edu/abs/2005Sci...310..834A}
}

@ARTICLE{Agol2021PSJ.....2....1A,
       author = {{Agol}, Eric and {Dorn}, Caroline and {Grimm}, Simon L. and {Turbet}, Martin and {Ducrot}, Elsa and {Delrez}, Laetitia and {Gillon}, Micha{\"e}l and {Demory}, Brice-Olivier and {Burdanov}, Artem and {Barkaoui}, Khalid and {Benkhaldoun}, Zouhair and {Bolmont}, Emeline and {Burgasser}, Adam and {Carey}, Sean and {de Wit}, Julien and {Fabrycky}, Daniel and {Foreman-Mackey}, Daniel and {Haldemann}, Jonas and {Hernandez}, David M. and {Ingalls}, James and {Jehin}, Emmanuel and {Langford}, Zachary and {Leconte}, J{\'e}r{\'e}my and {Lederer}, Susan M. and {Luger}, Rodrigo and {Malhotra}, Renu and {Meadows}, Victoria S. and {Morris}, Brett M. and {Pozuelos}, Francisco J. and {Queloz}, Didier and {Raymond}, Sean N. and {Selsis}, Franck and {Sestovic}, Marko and {Triaud}, Amaury H.~M.~J. and {Van Grootel}, Valerie},
        title = "{Refining the Transit-timing and Photometric Analysis of TRAPPIST-1: Masses, Radii, Densities, Dynamics, and Ephemerides}",
      journal = {\psj},
         year = 2021,
        month = feb,
       volume = {2},
       number = {1},
          eid = {1},
        pages = {1},
          doi = {10.3847/PSJ/abd022},
archivePrefix = {arXiv},
       eprint = {2010.01074},
 primaryClass = {astro-ph.EP},
       adsurl = {https://ui.adsabs.harvard.edu/abs/2021PSJ.....2....1A}
}

@ARTICLE{Melo2024,
       author = {{Melo}, Edypo and {Souto}, Diogo and {Cunha}, Katia and {Smith}, Verne V. and {Wanderley}, F{\'a}bio and {Grilo}, Vinicius and {Camara}, Deusalete and {Murta}, Kely and {Hejazi}, Neda and {Crossfield}, Ian J.~M. and {Teske}, Johanna and {Luque}, Rafael and {Zhang}, Michael and {Bean}, Jacob},
        title = "{Stellar Characterization and Chemical Abundances of Exoplanet-hosting M Dwarfs from APOGEE Spectra: Future JWST Targets}",
      journal = {\apj},
         year = 2024,
        month = oct,
       volume = {973},
       number = {2},
          eid = {90},
        pages = {90},
          doi = {10.3847/1538-4357/ad5004},
archivePrefix = {arXiv},
       eprint = {2406.00111},
 primaryClass = {astro-ph.SR},
       adsurl = {https://ui.adsabs.harvard.edu/abs/2024ApJ...973...90M}
}

@ARTICLE{Gaia2021,
       author = {{Gaia Collaboration} and {Brown}, A.~G.~A. and {Vallenari}, A. and {Prusti}, T. and {de Bruijne}, J.~H.~J. and {Babusiaux}, C. and {Biermann}, M. and {Creevey}, O.~L. and {Evans}, D.~W. and {Eyer}, L. and {Hutton}, A. and {Jansen}, F. and {Jordi}, C. and {Klioner}, S.~A. and {Lammers}, U. and {Lindegren}, L. and {Luri}, X. and {Mignard}, F. and {Panem}, C. and {Pourbaix}, D. and {Randich}, S. and {Sartoretti}, P. and {Soubiran}, C. and {Walton}, N.~A. and {Arenou}, F. and {Bailer-Jones}, C.~A.~L. and {Bastian}, U. and {Cropper}, M. and {Drimmel}, R. and {Katz}, D. and {Lattanzi}, M.~G. and {van Leeuwen}, F. and {Bakker}, J. and {Cacciari}, C. and {Casta{\~n}eda}, J. and {De Angeli}, F. and {Ducourant}, C. and {Fabricius}, C. and {Fouesneau}, M. and {Fr{\'e}mat}, Y. and {Guerra}, R. and {Guerrier}, A. and {Guiraud}, J. and {Jean-Antoine Piccolo}, A. and {Masana}, E. and {Messineo}, R. and {Mowlavi}, N. and {Nicolas}, C. and {Nienartowicz}, K. and {Pailler}, F. and {Panuzzo}, P. and {Riclet}, F. and {Roux}, W. and {Seabroke}, G.~M. and {Sordo}, R. and {Tanga}, P. and {Th{\'e}venin}, F. and {Gracia-Abril}, G. and {Portell}, J. and {Teyssier}, D. and {Altmann}, M. and {Andrae}, R. and {Bellas-Velidis}, I. and {Benson}, K. and {Berthier}, J. and {Blomme}, R. and {Brugaletta}, E. and {Burgess}, P.~W. and {Busso}, G. and {Carry}, B. and {Cellino}, A. and {Cheek}, N. and {Clementini}, G. and {Damerdji}, Y. and {Davidson}, M. and {Delchambre}, L. and {Dell'Oro}, A. and {Fern{\'a}ndez-Hern{\'a}ndez}, J. and {Galluccio}, L. and {Garc{\'\i}a-Lario}, P. and {Garcia-Reinaldos}, M. and {Gonz{\'a}lez-N{\'u}{\~n}ez}, J. and {Gosset}, E. and {Haigron}, R. and {Halbwachs}, J. -L. and {Hambly}, N.~C. and {Harrison}, D.~L. and {Hatzidimitriou}, D. and {Heiter}, U. and {Hern{\'a}ndez}, J. and {Hestroffer}, D. and {Hodgkin}, S.~T. and {Holl}, B. and {Jan{\ss}en}, K. and {Jevardat de Fombelle}, G. and {Jordan}, S. and {Krone-Martins}, A. and {Lanzafame}, A.~C. and {L{\"o}ffler}, W. and {Lorca}, A. and {Manteiga}, M. and {Marchal}, O. and {Marrese}, P.~M. and {Moitinho}, A. and {Mora}, A. and {Muinonen}, K. and {Osborne}, P. and {Pancino}, E. and {Pauwels}, T. and {Petit}, J. -M. and {Recio-Blanco}, A. and {Richards}, P.~J. and {Riello}, M. and {Rimoldini}, L. and {Robin}, A.~C. and {Roegiers}, T. and {Rybizki}, J. and {Sarro}, L.~M. and {Siopis}, C. and {Smith}, M. and {Sozzetti}, A. and {Ulla}, A. and {Utrilla}, E. and {van Leeuwen}, M. and {van Reeven}, W. and {Abbas}, U. and {Abreu Aramburu}, A. and {Accart}, S. and {Aerts}, C. and {Aguado}, J.~J. and {Ajaj}, M. and {Altavilla}, G. and {{\'A}lvarez}, M.~A. and {{\'A}lvarez Cid-Fuentes}, J. and {Alves}, J. and {Anderson}, R.~I. and {Anglada Varela}, E. and {Antoja}, T. and {Audard}, M. and {Baines}, D. and {Baker}, S.~G. and {Balaguer-N{\'u}{\~n}ez}, L. and {Balbinot}, E. and {Balog}, Z. and {Barache}, C. and {Barbato}, D. and {Barros}, M. and {Barstow}, M.~A. and {Bartolom{\'e}}, S. and {Bassilana}, J. -L. and {Bauchet}, N. and {Baudesson-Stella}, A. and {Becciani}, U. and {Bellazzini}, M. and {Bernet}, M. and {Bertone}, S. and {Bianchi}, L. and {Blanco-Cuaresma}, S. and {Boch}, T. and {Bombrun}, A. and {Bossini}, D. and {Bouquillon}, S. and {Bragaglia}, A. and {Bramante}, L. and {Breedt}, E. and {Bressan}, A. and {Brouillet}, N. and {Bucciarelli}, B. and {Burlacu}, A. and {Busonero}, D. and {Butkevich}, A.~G. and {Buzzi}, R. and {Caffau}, E. and {Cancelliere}, R. and {C{\'a}novas}, H. and {Cantat-Gaudin}, T. and {Carballo}, R. and {Carlucci}, T. and {Carnerero}, M.~I. and {Carrasco}, J.~M. and {Casamiquela}, L. and {Castellani}, M. and {Castro-Ginard}, A. and {Castro Sampol}, P. and {Chaoul}, L. and {Charlot}, P. and {Chemin}, L. and {Chiavassa}, A. and {Cioni}, M. -R.~L. and {Comoretto}, G. and {Cooper}, W.~J. and {Cornez}, T. and {Cowell}, S. and {Crifo}, F. and {Crosta}, M. and {Crowley}, C. and {Dafonte}, C. and {Dapergolas}, A. and {David}, M. and {David}, P.},
        title = "{Gaia Early Data Release 3. Summary of the contents and survey properties}",
      journal = {\aap},
         year = 2021,
        month = may,
       volume = {649},
          eid = {A1},
        pages = {A1},
          doi = {10.1051/0004-6361/202039657},
archivePrefix = {arXiv},
       eprint = {2012.01533},
 primaryClass = {astro-ph.GA},
       adsurl = {https://ui.adsabs.harvard.edu/abs/2021A&A...649A...1G}
}

@article{Souto2020,
doi = {10.3847/1538-4357/ab6d07},
url = {https://dx.doi.org/10.3847/1538-4357/ab6d07},
year = {2020},
month = {feb},
publisher = {The American Astronomical Society},
volume = {890},
number = {2},
pages = {133},
author = {Souto, Diogo and Cunha, Katia and Smith, Verne V. and Allende Prieto, C. and Burgasser, Adam and Covey, Kevin and García-Hernández, D. A. and Holtzman, Jon A. and Johnson, Jennifer A. and Jönsson, Henrik and Mahadevan, Suvrath and Majewski, Steven R. and Masseron, Thomas and Shetrone, Matthew and Rojas-Ayala, Bárbara and Sobeck, Jennifer and Stassun, Keivan G. and Terrien, Ryan and Teske, Johanna and Wanderley, Fábio and Zamora, Olga},
title = {Stellar Characterization of M Dwarfs from the APOGEE Survey: A Calibrator Sample for M-dwarf Metallicities},
journal = {The Astrophysical Journal}
}

@INPROCEEDINGS{Wilson2010,
       author = {{Wilson}, John C. and {Hearty}, Fred and {Skrutskie}, Michael F. and {Majewski}, Steven and {Schiavon}, Ricardo and {Eisenstein}, Daniel and {Gunn}, Jim and {Blank}, Basil and {Henderson}, Chuck and {Smee}, Stephen and {Barkhouser}, Robert and {Harding}, Al and {Fitzgerald}, Greg and {Stolberg}, Todd and {Arns}, Jim and {Nelson}, Matt and {Brunner}, Sophia and {Burton}, Adam and {Walker}, Eric and {Lam}, Charles and {Maseman}, Paul and {Barr}, Jim and {Leger}, French and {Carey}, Larry and {MacDonald}, Nick and {Horne}, Todd and {Young}, Erick and {Rieke}, George and {Rieke}, Marcia and {O'Brien}, Tom and {Hope}, Steve and {Krakula}, John and {Crane}, Jeff and {Zhao}, Bo and {Carr}, Mike and {Harrison}, Craig and {Stoll}, Robert and {Vernieri}, Mary A. and {Holtzman}, Jon and {Shetrone}, Matt and {Allende-Prieto}, Carlos and {Johnson}, Jennifer and {Frinchaboy}, Peter and {Zasowski}, Gail and {Bizyaev}, Dmitry and {Gillespie}, Bruce and {Weinberg}, David},
        title = "{The Apache Point Observatory Galactic Evolution Experiment (APOGEE) high-resolution near-infrared multi-object fiber spectrograph}",
    booktitle = {Ground-based and Airborne Instrumentation for Astronomy III},
         year = 2010,
       editor = {{McLean}, Ian S. and {Ramsay}, Suzanne K. and {Takami}, Hideki},
       series = {Society of Photo-Optical Instrumentation Engineers (SPIE) Conference Series},
       volume = {7735},
        month = jul,
          eid = {77351C},
        pages = {77351C},
          doi = {10.1117/12.856708},
       adsurl = {https://ui.adsabs.harvard.edu/abs/2010SPIE.7735E..1CW}
}

@ARTICLE{Gustafsson2008,
       author = {{Gustafsson}, B. and {Edvardsson}, B. and {Eriksson}, K. and {J{\o}rgensen}, U.~G. and {Nordlund}, {\r{A}}. and {Plez}, B.},
        title = "{A grid of MARCS model atmospheres for late-type stars. I. Methods and general properties}",
      journal = {\aap},
         year = 2008,
        month = aug,
       volume = {486},
       number = {3},
        pages = {951-970},
          doi = {10.1051/0004-6361:200809724},
archivePrefix = {arXiv},
       eprint = {0805.0554},
 primaryClass = {astro-ph},
       adsurl = {https://ui.adsabs.harvard.edu/abs/2008A&A...486..951G}
}

@ARTICLE{Souto2017,
   author = {{Souto}, D. and {Cunha}, K. and {Garc{\'{\i}}a-Hern{\'a}ndez}, D.~A. and 
	{Zamora}, O. and {Allende Prieto}, C. and {Smith}, V.~V. and 
	{Mahadevan}, S. and {Blake}, C. and {Johnson}, J.~A. and {J{\"o}nsson}, H. and 
	{Pinsonneault}, M. and {Holtzman}, J. and {Majewski}, S.~R. and 
	{Shetrone}, M. and {Teske}, J. and {Nidever}, D. and {Schiavon}, R. and 
	{Sobeck}, J. and {Garc{\'{\i}}a P{\'e}rez}, A.~E. and {G{\'o}mez Maqueo Chew}, Y. and 
	{Stassun}, K.},
    title = "{Chemical Abundances of M-dwarfs from the APOGEE Survey. I. The Exoplanet Hosting Stars Kepler-138 and Kepler-186}",
  journal = {\apj},
archivePrefix = "arXiv",
   eprint = {1612.01598},
 primaryClass = "astro-ph.SR",
     year = 2017,
    month = feb,
   volume = 835,
      eid = {239},
    pages = {239},
      doi = {10.3847/1538-4357/835/2/239},
   adsurl = {https://ui.adsabs.harvard.edu/abs/2017ApJ...835..239S}
}

@MISC{Masseron2016,
   author = {{Masseron}, T. and {Merle}, T. and {Hawkins}, K.},
    title = "{BACCHUS: Brussels Automatic Code for Characterizing High accUracy Spectra}",
howpublished = {Astrophysics Source Code Library},
     year = 2016,
    month = may,
archivePrefix = "ascl",
   eprint = {1605.004},
      doi = {10.20356/C4TG6R},
   adsurl = {https://ui.adsabs.harvard.edu/abs/2016ascl.soft05004M}
}

@ARTICLE{AlvarezPLez1998,
   author = {{Alvarez}, R. and {Plez}, B.},
    title = "{Near-infrared narrow-band photometry of M-giant and Mira stars: models meet observations}",
  journal = {\aap},
   eprint = {astro-ph/9710157},
     year = 1998,
    month = feb,
   volume = 330,
    pages = {1109-1119},
   adsurl = {https://ui.adsabs.harvard.edu/abs/1998A%26A...330.1109A},
doi = {10.48550/arXiv.astro-ph/9711225}
}

@MISC{Plez2012,
   author = {{Plez}, B.},
    title = "{Turbospectrum: Code for spectral synthesis}",
howpublished = {Astrophysics Source Code Library},
     year = 2012,
    month = may,
archivePrefix = "ascl",
   eprint = {1205.004},
   adsurl = {https://ui.adsabs.harvard.edu/abs/2012ascl.soft05004P}
}
\bibliographystyle{aasjournalv7}

%% This command is needed to show the entire author+affiliation list when
%% the collaboration and author truncation commands are used.  It has to
%% go at the end of the manuscript.
%\allauthors

%% Include this line if you are using the \added, \replaced, \deleted
%% commands to see a summary list of all changes at the end of the article.
%\listofchanges

\end{document}